\newcounter{myctr}
\def\myitem{\refstepcounter{myctr}\bibfont\noindent\ifnum\themyctr>9\else\phantom{0}\fi\hangindent17pt\themyctr.\enskip}

\documentclass[reprint,showpacs,aps,pra,onecolumn]{revtex4-1}
\usepackage{amssymb,amsmath}
\usepackage{mathtools}
\usepackage[ruled]{algorithm2e}
\usepackage{subfigure}

\def\ket#1{|#1 \rangle}
\def\ketbra#1#2{|#1 \rangle \langle #2|}

\def\kmax{k_{\textrm{max}}}
\def\gammac{\gamma_{\textrm{classical}}}
\def\alphac{\alpha_{\textrm{classical}}}
\def\rhoini{\rho_{\textrm{ini}}}
\def\rhoun{\rho_{\textrm{unknown}}}

\begin{document}

\markboth{Cohn \it{et al.}}
{GROVER'S SEARCH WITH LOCAL AND TOTAL DEPOLARIZING CHANNEL ERRORS}


\title{GROVER'S SEARCH WITH LOCAL AND TOTAL DEPOLARIZING\\CHANNEL ERRORS: COMPLEXITY ANALYSIS}

\author{ILAN COHN, ANDR\'E L. FONSECA DE OLIVEIRA and EFRAIN BUKSMAN}
\address{Facultad de Ingenier\'ia, Universidad ORT Uruguay, Cuareim 1451\\
Montevideo, CP 11100, Uruguay\\
icohn@uni.ort.edu.uy \\ fonseca@ort.edu.uy \\ buksman@ort.edu.uy}

%

\author{JES\'US GARC\'IA L\'OPEZ DE LACALLE}
\address{Escuela T\'ecnica Superior de Ingenier\'ia de Sistemas Inform\'aticos\\ Universidad Polit\'ecnica de Madrid\\Calle Alan Turing s/n, 28031 Madrid, Spain\\
jglopez@etsisi.upm.es}

\begin{abstract}
In this article the effect of noise on Grover's algorithm is analyzed, modeled as a total depolarizing channel (TDCh) and a local depolarizing channel in each qubit (LDCh). The focus was not in error correction (e.g. by the fault-tolerant method), but to provide an insight to the kind of error, or degradation, that needs to be corrected. In the last years analytical results regarding mainly the TDCh model have been obtained. In this paper we extend these previous results to the local case, concluding that the degradation of Grover's algorithm with the latter is worse than the former. It has been shown that for both cases with an $N$-dependent small enough error-width, smaller than $1/\sqrt{N}$ for total error and $1/(\sqrt{N}\log_2{N})$ for the local case, correction is not needed.
\end{abstract}

\maketitle

\section{Introduction}
\label{sec_introduction}

Quantum systems cannot be completely isolated from the environment. When a quantum system is to be controlled externally, e.g. gate applications, state preparations and others, additional errors are generated. Such effect on quantum algorithms has been studied lately by several authors \cite{Azuma_2002,Shapira_2003,Salas_2008,Gawron_2012,Ambainis_2013_Grover}.

While there is a fault-tolerant model of quantum computing based on the correction of errors below a certain threshold \cite{Aharonov_1997,Gottesman_1998}, this method is very expensive in computational resources. Therefore, it is important to diagnose the performance of quantum algorithms affected by noise before making any correction.

Grover's quantum algorithm is an oracle-based search of an unordered database that, without noise, presents a quadratic speedup over the brute-force classical search algorithm \cite{Grover_1996,Grover_1997,Grover_2001}. However, this advantage is affected drastically if the oracle is faulty \cite{Regev_2008,Temme_2014} or in the presence of noise \cite{Chen_2003}. 

Local error models are important due to the fact that any implementation of a quantum circuit are affected by gate errors. In this article we study the effect of a known rate of noise on Grover's search algorithm, extending the work done in \cite{Vrana_2014} with total depolarizing channel error model (TDCh), and also analyzing the effects of local depolarizing channel errors (LDCh). For both model errors, it is shown that quantum speedup disappears when the error rate is constant. Furthermore, we have concluded that the degradation of Grover's algorithm with local error (LDCh) is worse than with total error (TDCh). This is coherent with recent results \cite{Demkowicz_2015}.

The article is divided as follows: in order to introduce the equations used later, in section 2 Grover's algorithm is briefly explained. Sections 3 and 4 deal with the effect of total and local depolarizing error (TDCh and LDCh) on Grover's search, respectively. Some conclusions are drawn in section 5. The major calculus are given in the appendixes.

\section{Grover's quantum search algorithm}
\label{sec_grover_intro}

Grover's quantum search algorithm is known to be optimal, in some sense, to solve the 
problem of finding a marked element in a unsorted database of $N$ elements using 
$k_{Gr} = \left\lfloor \frac{\pi}{4}\sqrt{N}\right\rfloor $ oracle queries 
\cite{Zalka_1999}. 
Such problem may be shortly stated as follows: suppose we have a database of $N=2^{n}$
of quantum states, being $n$ the number of qubits that span the Hilbert space (
$\mathcal{H}=\mathbb{C}^{2^n}$), and an unknown marked state 
among them. Given an \emph{oracle} ( or \emph{black box} ) to identify whether an 
element is the one being searched, the goal is to find such marked state with 
high probability and in as few steps as possible.

%
%
Let $\ket{t}$ denote the target basis state and $\ket{s} = 
\frac{1}{\sqrt{N}} \sum_{i=0}^{N-1} \ket{i}$ the superposition of all basis states. 
To be able to introduce quantum error, we use the density matrix notation ($\mathcal{H}=\mathbb{C}^{2^n}\times\mathbb{C}^{2^n}$), obtaining: $\rho_{t}=\ketbra{t}{t}$ and  
$\rho_{0}=\ketbra{s}{s}$, respectively.
The algorithm is represented in {\bf Algorithm \ref{alg_grover}}.

\begin{algorithm}[h]
\BlankLine
\label{alg_grover}
\caption{Grover's search algorithm}
{\bf 1.} Set up the superposition state $\rho_{0}$.\\
{\bf 2.} Apply the oracle operator $O=2\left| t \right\rangle \left\langle t\right|-I$.\\
{\bf 3.} Apply the diffusion operator $D=2\left|s\right\rangle \left\langle s\right|-I$.\\
{\bf 4.} Repeat steps 2 and 3 $\left\lfloor \frac{\pi}{4}\sqrt{N}\right\rfloor -1 $
times.\\
{\bf 5.} Perform measurements in the canonical basis in each qubit. The target state will emerge with high probability as $N\gg1$.\\
\BlankLine
\end{algorithm}

\medskip
It is straightforward that after applying the oracle operator and diffusion operator $k$ times, the result is

\begin{equation}
\label{eq_rho_k_groverOp}
	\rho(k)=G^{k}\rho_{0}(G^{\dagger})^{k},
\end{equation}
where $G=D O$ is also known as the Grover operator. It can be shown that the density operator obtained afterwards is
\begin{equation} 
\label{eq_rho_k_sk}
	\rho(k)  =  \ketbra{s_{k}}{s_{k}},
\end{equation}
where 

\begin{eqnarray} \label{eq_rho_k}
	\ket{s_{k}} &=&  \sin\left( (2 k+1)\theta\right)\ket{t} +\cos((2 k+1)\theta)\ket{\bar{t}}, \nonumber \\ 
	\ket{\bar{t}} &=&  \frac{1}{\sqrt{N-1}}  \sum_{\substack{i=0 \\ i\neq t}}^{N-1} \ket{i} \ \, \text{and} \ \,
	\theta  =  \arcsin\left(\frac{1}{\sqrt{N}}\right).
\end{eqnarray}

Hence, the probability of success (measuring the marked element) after $k$ steps is
\begin{equation}\label{eq_p_k}
	p(k) = \sin^{2}\left(\left(2k+1\right)\theta\right).
\end{equation}

\section{Grover's search algorithm with Total Depolarizing Channel}
\label{sec_groverTDC}

In this section we analyze how the error modeled as the Total Depolarizing Channel 
(TDCh) affects Grover's search algorithm. Similar work, but with different focus, has 
been done by Vrana et al \cite{Vrana_2014}. Instead of minimizing the mean cost as 
done in \cite{Vrana_2014} we propose a maximization of the probability, because a 
closed form  expression can be obtained. This enables us to compare the results with the error modeled as the Local Depolarizing Channel (LDCh), in Sec. \ref{sec_groverLDC}. The TDCh is an error model, which maps an $n$-qubit state $\rho$ to the maximally mixed state $\frac{I}{N}$ with probability $\gamma$, and leaves it unchanged with probability $(1-\gamma)$ \cite{Nielsen_2000}, given by
\begin{equation} 
\label{eq_dep_channel_total}
	\varepsilon(\rho, \gamma) = (1-\gamma)\rho + \gamma \frac{I}{N}.
\end{equation}
This can be interpreted in the probabilistic sense as a Binomial distribution, where \emph{success} or \emph{failure} should be interpreted as having $\rho$ or $\frac{I}{N}$ respectively. We will use these terms indistinctly.

\subsection{State evolution with error}
\label{subsec_groverTDC_State}

We now consider the effects of the TDCh in Grover's algorithm. That is, every step of the algorithm involves applying the Grover operator, followed by the TDCh error. Since the TDCh error commutes with any unitary operator, the density matrix obtained after $k$ steps is
\begin{equation} 
\label{eq_rho_k_techo_total}
	\hat{\rho}(k,\gamma) = (1-\gamma)^k \rho(k) + (1 - (1-\gamma)^k)\frac{I}{N},
\end{equation}
where $\rho(k)$ is given by Eq. (\ref{eq_rho_k}). This has a very straightforward interpretation: we have probability of success $(1-\gamma)^k$ and probability of failure  $1 - (1-\gamma)^k$. 

This implies the probability of finding the marked element is 
\begin{equation} 
\label{eq_p_k_techo_total}
	\hat{p}(k,\gamma) = (1-\gamma)^{k} p(k) + \frac{1-(1-\gamma)^k}{N},
\end{equation}
where $p(k)$ is given by Eq. (\ref{eq_p_k}). Notice how the maximally mixed state $\frac{I}{N}$ introduces uniform probability over all possible states.

\begin{figure} [htbp]
\centering
\includegraphics[width=0.65\textwidth]{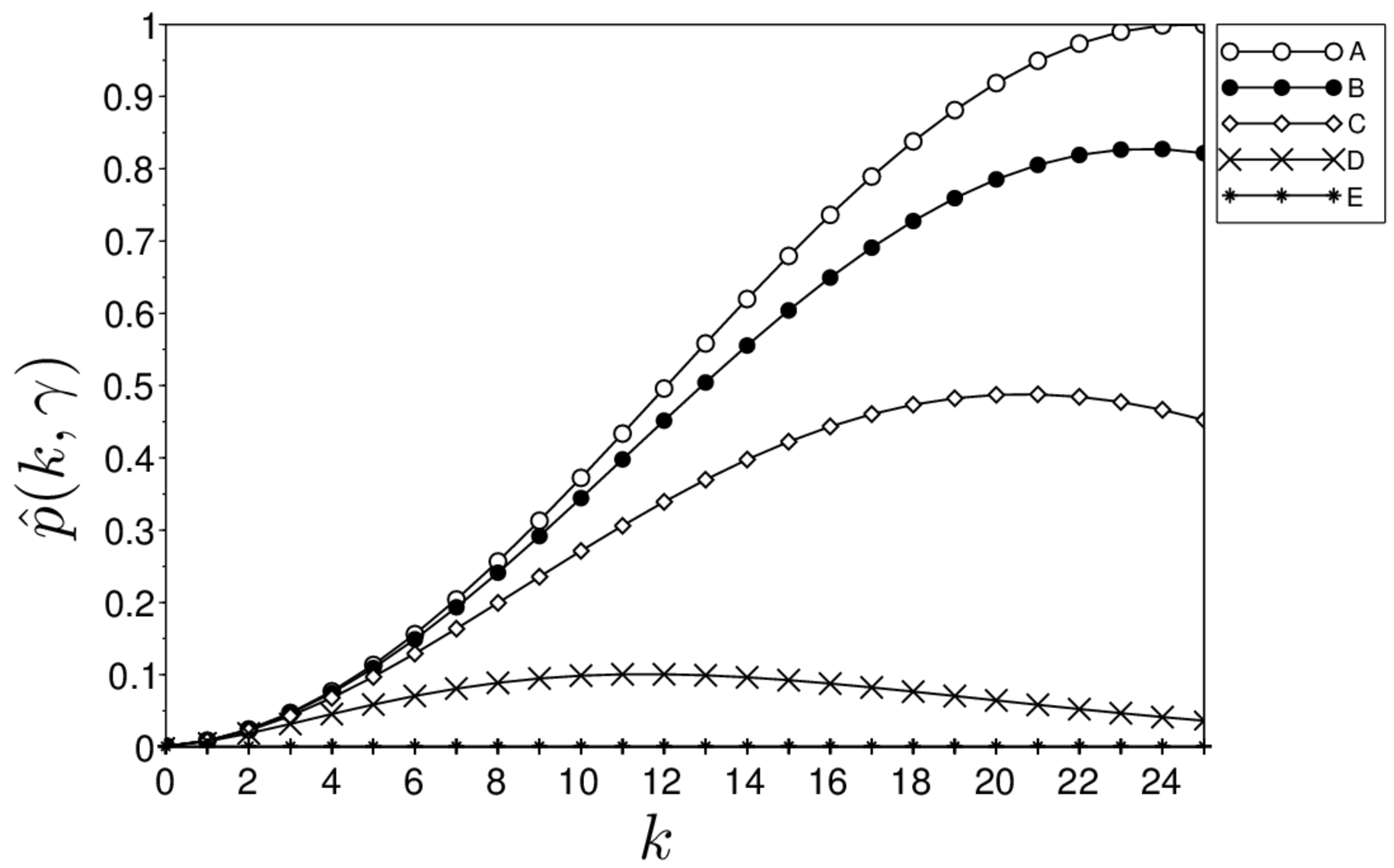}
\caption{Probability vs. number of steps with $n = 10$ qubits and different values of $\gamma$: $0$ (A), $1/(4\sqrt{N})$ (B), $1/\sqrt{N}$ (C), $4/\sqrt{N}$ (D) and 1 (E).} 
\label{fig_pk_vs_k_Nfijo_familiaGam}
\end{figure}

\begin{figure}[htbp]
\centering
\includegraphics[width=0.65\textwidth]{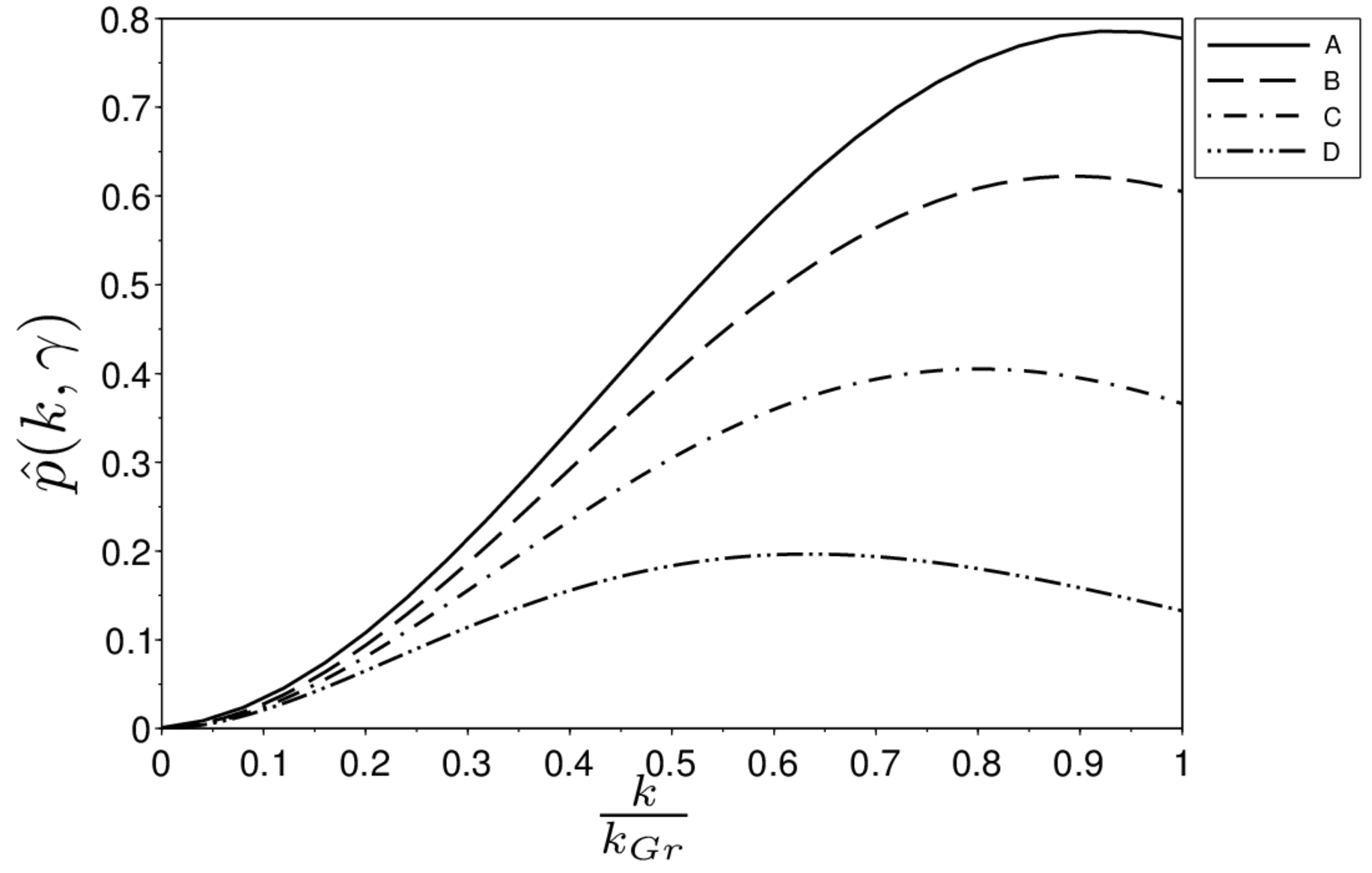} 
\caption{Probability vs. normalized number of steps with $\gamma = 0.01$ and different number of qubits $n$: $10$ (A), $12$ (B), $14$ (C) and $16$ (D).}
\label{fig_pk_vs_kNorm_GamFijo_familiaN}
\end{figure}

\begin{figure}[!htb]
\centering
\includegraphics[width=0.65\textwidth]{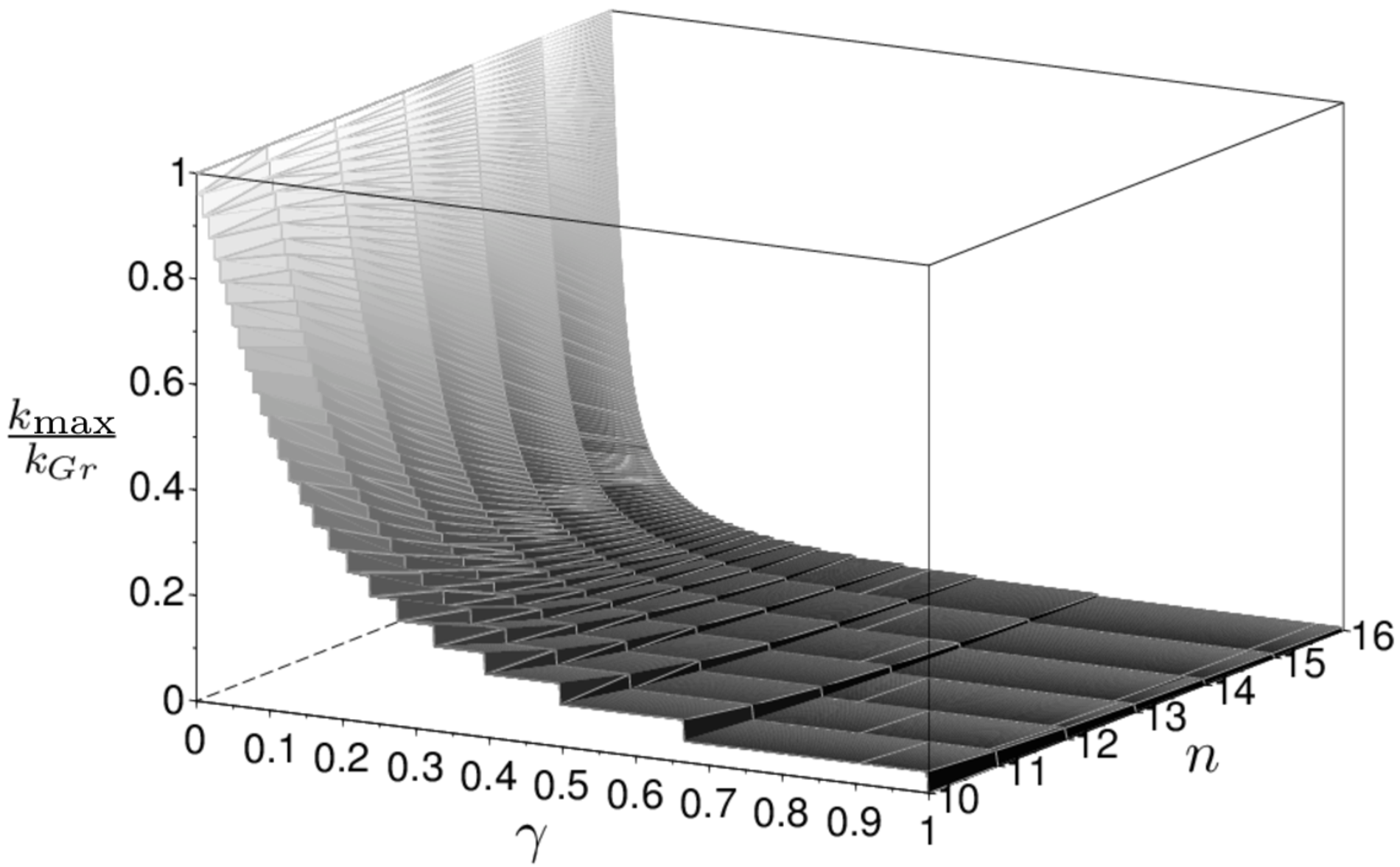} 
\caption{Number of steps $\kmax/k_{Gr}$ vs. width of noise $\gamma$ vs. number of qubits $n$. }
\label{fig_sec_groverTDC_subsec_modeloError_3}
\end{figure}

 As can be seen from Fig. \ref{fig_pk_vs_k_Nfijo_familiaGam} and Fig. 
\ref{fig_pk_vs_kNorm_GamFijo_familiaN}, the first maxima of the probability 
$\hat{p}(k)$, at step $\kmax$, moves left with respect to $k_{Gr}$ as either $\gamma$ or $N$ increase \cite{Salas_2008}. This can also be appreciated on Fig. \ref{fig_sec_groverTDC_subsec_modeloError_3}, which also shows that for any number of qubits, e.g. $n = 10$, the behavior is similar. A good approximation of $\kmax$ can be deduced by maximizing Eq. (\ref{eq_p_k_techo_total}) (see \ref{sec_appendix_subsec_finding_kmax}), resulting in
\begin{equation}
\label{eq_k_max_teorico}
\kmax(\gamma) = \max\left( \left\lfloor  \frac{\pi - \arcsin{\delta} - \arcsin{\left(  \left[ 1-\frac{2}{N} \right] \delta \right) }}{4\theta}  \right\rfloor , 1\right)
\end{equation}
where
\begin{equation} 
\delta = \sqrt{\frac{1}{1 + {\left(\frac{4 \theta}{\ln{(1-\gamma)}}\right)} ^2}} \label{eq_k_max_teorico_delta}
\end{equation}
and $\theta$ is given by Eq. (\ref{eq_rho_k}).
We provide two approximations of Eq. (\ref{eq_k_max_teorico}): for $ n \gg 1$ $\left( N=2^n\right)$ and $0  \leq \gamma \ll \frac{2\pi}{\sqrt{N}}$
\begin{equation} 
\label{eq_k_max_taylor_gammachico}
	\kmax(\gamma) \approx    \left\lfloor \frac{\pi \sqrt{N}}{4} - \frac{N\gamma}{8} \right\rfloor ,
\end{equation}
and for $n \gg 1$ with $1 \geq \gamma \gg \frac{8}{\pi\sqrt{N}}$
\begin{equation} 
\label{eq_k_max_taylor_gammagrande}
	\kmax (\gamma) \approx  \max\left( \left\lfloor - \frac{1}{\ln{(1-\gamma)}} + \frac{1}{2} \sqrt{ 1 + \left( \frac{2}{\ln{(1-\gamma)}} \right)^2 } \right\rfloor, 1\right).
\end{equation}

\begin{figure}[!htb]
\centering
\includegraphics[width=0.65\textwidth]{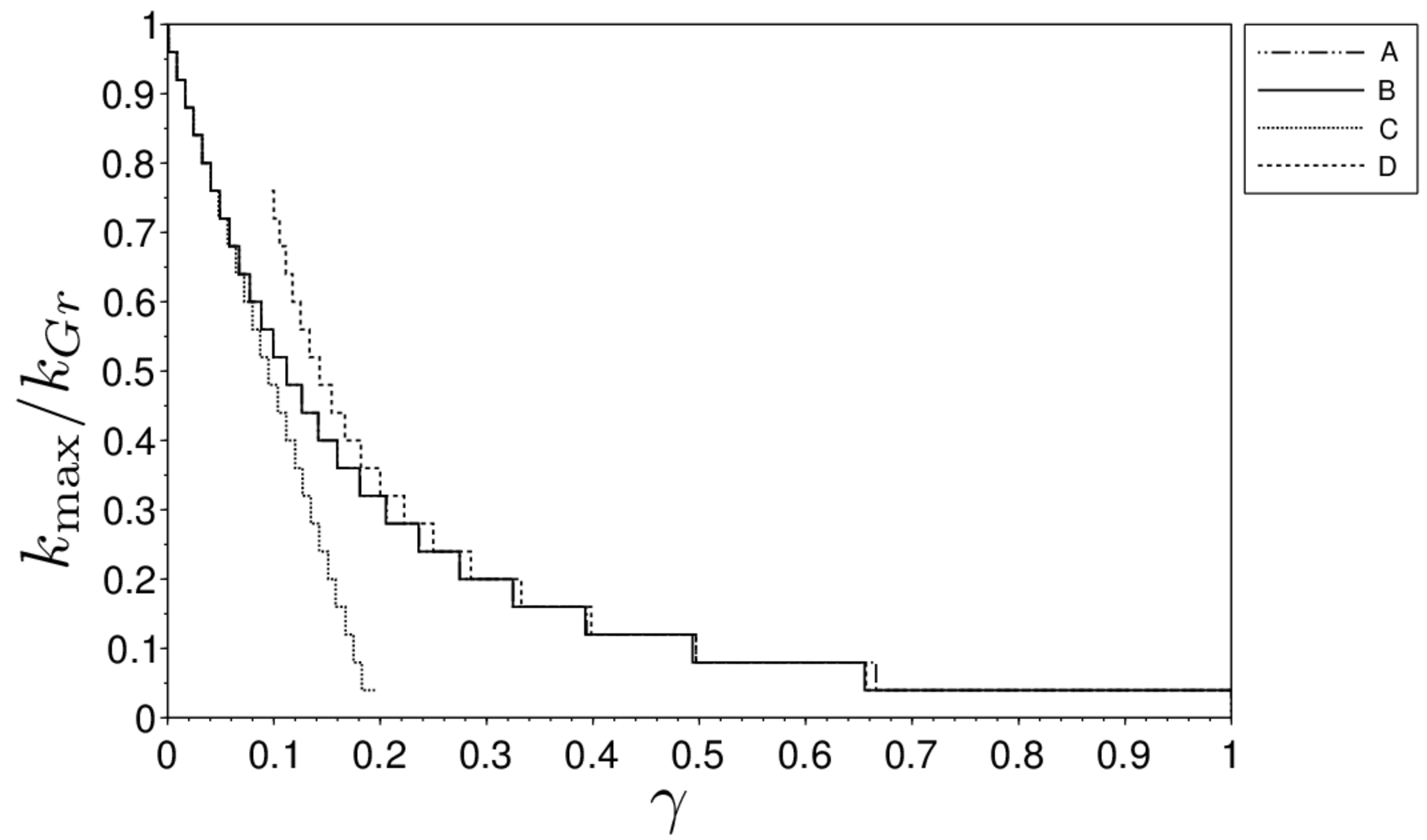}
\caption{Step of maximum probability vs. width of noise $\gamma$, with $n = 10$ qubits and different approximations of $\kmax$: obtained by simulation (A) and as in equations (\ref{eq_k_max_teorico}) (B), (\ref{eq_k_max_taylor_gammachico}) (C) and (\ref{eq_k_max_taylor_gammagrande}) (D).} 
\label{fig_sec_groverTDC_subsec_modeloError_5}
\end{figure}

Notice how for Eq. (\ref{eq_k_max_taylor_gammachico}), the approximation is valid only for small values of $\gamma$; while the approximation given by Eq. (\ref{eq_k_max_taylor_gammagrande}) is valid for a large range of $\gamma$. Fig. \ref{fig_sec_groverTDC_subsec_modeloError_5} shows different approximations of $\kmax$.

%
%
%
%
\subsection{Order estimation of the algorithm}
\label{subsec_groverTDC_orden}

If we apply classical amplification and stop at step $k$, we have the order of the algorithm given by the estimate mean cost (expected number of oracle queries): 
\begin{equation} \label{eq_costo_medio}
	MC_{k,\gamma} \coloneqq MC[k, \hat{p}(k, \gamma)] = \frac{k}{\hat{p}(k, \gamma)}.
\end{equation}

It is well-known that the classical brute-force search algorithm has order $\Theta(N)$ with a constant factor of $1/2$.

%
%
\subsubsection{Stopping the algorithm at $k_{Gr}$}
\label{subsubsec_groverTDC_orden_kGr}

If we were to stop the algorithm at $k_{Gr}$, we would have the following estimate cost for the algorithm:
\begin{equation} \label{eq_orden_kgr_gral}
	MC_{k_{Gr},\gamma} = \frac{k_{Gr} } { (1-\gamma)^{k_{Gr}}p\left(k_{Gr}\right) + \frac{1}{N} \left( 1-(1-\gamma)^{k_{Gr}} \right) }.
\end{equation}
Since we consider $N \gg 1$, we have: $p\left(k_{Gr}\right) \approx 1$ and $k_{Gr} = \left\lfloor \frac{\pi \sqrt{N}}{4} \right\rfloor \approx  \frac{\pi \sqrt{N}}{4} $, leaving Eq. (\ref{eq_orden_kgr_gral}) as: 
\begin{equation} 
\label{eq_orden_kgr_N_grande}
	MC_{k_{Gr},\gamma} \approx \frac{\frac{\pi \sqrt{N}}{4} } { (1-\gamma)^{ \frac{\pi \sqrt{N}}{4} } + \frac{1}{N} \left( 1-(1-\gamma)^{ \frac{\pi \sqrt{N}}{4} } \right)}.
\end{equation}

Notice that for any fixed $\gamma > 0$ the order is $\Theta(N \sqrt{N})$. For small values of $\gamma$ ($\frac{\pi \sqrt{N}}{4}\gamma \ll 1$) Eq. (\ref{eq_orden_kgr_gral}) becomes
\begin{equation} \label{eq_orden_kgr_N_grande_gamma_chico}
	MC_{k_{Gr},\gamma} \approx \frac{\pi\sqrt{N}}{4} \left( 1 + \frac{\pi\sqrt{N}}{4}\gamma \right).
\end{equation}

It is interesting to find the width of noise that keeps the quantum speedup (order of the algorithm $\Theta(\sqrt{N})$). Hence, it is straightforward that $\gamma$ needs to be a function of $N$. In fact, one needs $\gamma \ll 1/k_{Gr}$, as shown in \cite{Vrana_2014}.

However, Grover's quantum search algorithm can still outperform the brute-force classical search algorithm. In fact, there is a maximum width of noise ($\gammac$) that leaves the mean cost of Grover's quantum search algorithm $\frac{N}{2}$. Such is found by using Eq. (\ref{eq_orden_kgr_N_grande}):
\begin{equation} \label{eq_orden_kgr_gamma_clasico}
	\gammac \approx 1 - {\left(\frac{-1 + 2 k_{Gr}}{-1 + N}\right)}^{\frac{1}{k_{Gr}}},
\end{equation}
which for $N \gg 1$, becomes
\begin{equation} \label{eq_orden_kgr_gamma_clasico_N_grande}
	\gammac \approx \frac{4 \ln{\left( \frac{2\sqrt{N}}{\pi}\right)}}{\pi \sqrt{N}}.
\end{equation}

%
%
\subsubsection{Stopping the algorithm at $\kmax$}
\label{subsubsec_groverTDC_orden_kMax}

If we were to stop the algorithm at $\kmax(\gamma)$, Eq. (\ref{eq_costo_medio}) becomes $MC\left[\kmax(\gamma), \hat{p}\left( \kmax(\gamma), \gamma \right) \right]$. Therefore, in contrast to stopping the algorithm at $k_{Gr}$, for any fixed $\gamma > 0$ the algorithm is of order $\Theta(N)$. For large values of $N$ ($N \gg 1$) and small values of $\gamma$ ($\sqrt{N}\gamma/(2\pi) \ll 1$), the mean cost becomes
\begin{equation} 
\label{eq_orden_kmax_N_grande_gamma_chico}
	MC_{\kmax(\gamma), \gamma} \approx \frac{\pi\sqrt{N}}{4} \left( 1 + \frac{\pi\sqrt{N}}{4}\gamma\left( 1 - \frac{2}{\pi^2}  \right) \right).
\end{equation}
Comparing this result with Eq. (\ref{eq_orden_kgr_N_grande_gamma_chico}), we can see that an extra constant factor of approximately $0.80$ appears. Then, we need $\gamma \ll 1/(0.8k_{Gr})$ to mantain the quadratic speedup (order of the algorithm $\Theta(\sqrt{N})$).

For any fixed $\gamma > 0$, the mean cost becomes
\begin{equation} 
\label{eq_orden_kmax_gama_grande}
	MC_{\kmax(\gamma), \gamma} \approx \frac{N}{9-8\gamma}.
\end{equation}
As a result, the maximum width of noise that leaves $MC\left[\kmax(\gamma), \hat{p}(\kmax(\gamma), \gamma) \right]$ classical ($N/2$) is independent of $N$, i.e. $\gammac \approx 7/8$.

\section{Grover's search algorithm with Local Depolarizing Channel}
\label{sec_groverLDC}

The Local Depolarizing Channel (LDCh), maps an $n$-qubit state $\rho$ to a mixed state applying the depolarizing channel in every qubit independently, i.e.
\begin{equation} 
\label{LDCh}
	\varepsilon(\rho)=\varepsilon_1(\rho,\alpha)\circ\varepsilon_2(\rho,\alpha) \circ \dots \circ \varepsilon_n(\rho,\alpha),	
\end{equation}
where $\varepsilon_i(\rho,\alpha)$ corresponds to the depolarizing channel (Eq. (\ref{eq_dep_channel_total})) acting in the qubit $i$, and $\alpha$ the probability of error. Analogously to Sec. \ref{sec_groverTDC}, we define the probability of obtaining the marked element after $k$ steps as $\hat{p}^L(k, \alpha)$, and the step at which the first maxima of the probability occurs as $\kmax^L(\alpha)$.

\subsection{Operator-sum representation for Grover state evolution with LDCh} 
\label{subsec_groverLDC_kraus}

For one qubit states the Depolarizing Channel error (Eq. (\ref{eq_dep_channel_total})) can be expressed using Kraus' operators (Operator-sum representation), as
\begin{equation} 
\label{eq_dep_channel_1qb}
	\varepsilon(\rho, \alpha) = \sum_{l=0}^{4}{M_l \rho M_l^\dagger},
\end{equation}
where $M_0 = \sqrt{\left( 1 - \frac{3\alpha}{4} \right)} \sigma_0$ and $M_i = \sqrt{\frac{\alpha}{4}} \sigma_i$ ($i \in \{1,2,3\}$), being $\sigma_\mu$ the Pauli matrices.

In the case applying local errors in all qubits in a $n$-qubits state there are $4^n$ operators (all posible combinations), which are of the form \cite{Nielsen_2000}

\begin{equation} \label{eq_kraus_operators_LDCh}
	M_c = \sqrt{\left( 1 - \frac{3\alpha}{4} \right)^m \left( \frac{\alpha}{4}\right)^{n-m}} \bigotimes_{q=1}^n \sigma_{i_q},  
\end{equation}
where $\sigma_{i_q}$ is the $i_q$ Pauli matrix ($i_q \in {0, \ldots, 3}$) and $m$ is the amount of $\sigma_0$ ($I$) operators present in a particular $M_c$. 

If we consider one step of the algorithm as applying the Grover operator $G$ and the LDCh, we get the following state after $k$ steps
\begin{equation} \label{eq_ro_k_techo_LDCh}
	\hat{\rho}^L(k, \alpha) = \sum_{c_1}^{4^n} \cdots \sum_{c_k}^{4^n} \left[ O_{\Pi} \,  \rhoini  \, O_{\Pi}^\dagger\right], 
\end{equation}
where
\begin{equation} \label{eq_ro_k_techo_LDCh_parte}
O_{\Pi} = \prod_{l=1}^k\left( M_{c_l} G\right),
\end{equation}
$l$ represents the step, each $M_{c_l}$ has the form of Eq. (\ref{eq_kraus_operators_LDCh}) and $\rhoini=\ketbra{s}{s}$ is the initial Grover state.

For the first step ($k=1$), such probability is
\begin{eqnarray} 
	\hat{p}^L(1,\alpha) & = & \frac{1}{2^{3 n -4}} \left[ 2^n(2^{n-1}-1)\left(1-\frac{\alpha}{2} \right)^n + \right. \nonumber \\
& & + \left. \frac{1}{16}\left(2^n-4 \right)^2 \right],
\label{eq_p1_alpha_exacto}
\end{eqnarray} 
which for $n\gg1$ becomes
\begin{equation} 
\label{eq_p1_alpha_n_grande}
	\hat{p}^L(1,\alpha) \approx \frac{1}{2^{n}} \left[ 2^3 \left(1-\frac{\alpha}{2} \right)^n +1 \right].
\end{equation} 

Due to the exponentially increasing complexity of Eq. (\ref{eq_ro_k_techo_LDCh}), we propose to analyze $\hat{p}^L(k, \alpha)$ using lower and upper bounds in terms of the TDCh model (because of simplicity and relatively tight bounds), and a first order approximation.

\subsection{Probability Bounds}
\label{subsec_groverLDC_ProbBounds}

\begin{figure} [!htbp]
\centering
\includegraphics[width=0.45\textwidth,keepaspectratio=true]{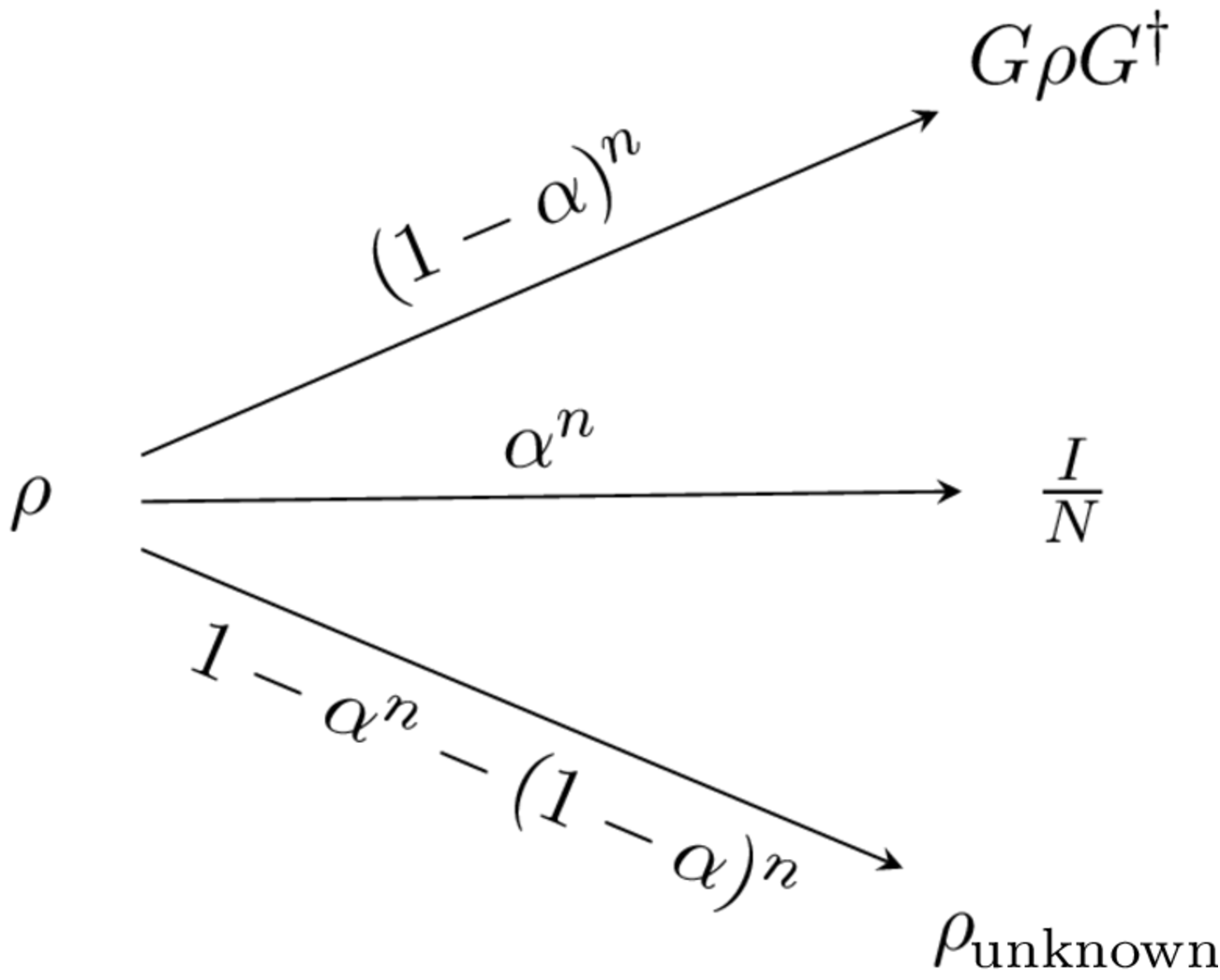}
 \caption{Ternary tree at step $k$ of the algorithm with LDCh.} 
\label{fig_subsec_groverLDC_cota_arbol}
\end{figure}

Grover's algorithm with LDCh error can be interpreted as a ternary tree where each depth-level corresponds to a step. The initial node is given by $\rhoini = \ketbra{s}{s}$, and in each node we have the paths shown in Fig. \ref{fig_subsec_groverLDC_cota_arbol}. In fact, the actual probability of getting the marked element at a step $k$ is obtained by taking all the possible paths (which are $4^{nk}$ if one uses the Kraus' operators).

\begin{figure}[htpb]
	\centering
	\subfigure[Probability evaluated at $k_{Gr}$ vs. width of error $\alpha$ for $n = 8$ qubits and its bounds.]{
	\label{cotas_nuevas_pk_vs_alfa_kGr_NFijo} 
	\includegraphics[width=0.65\textwidth,keepaspectratio=true]{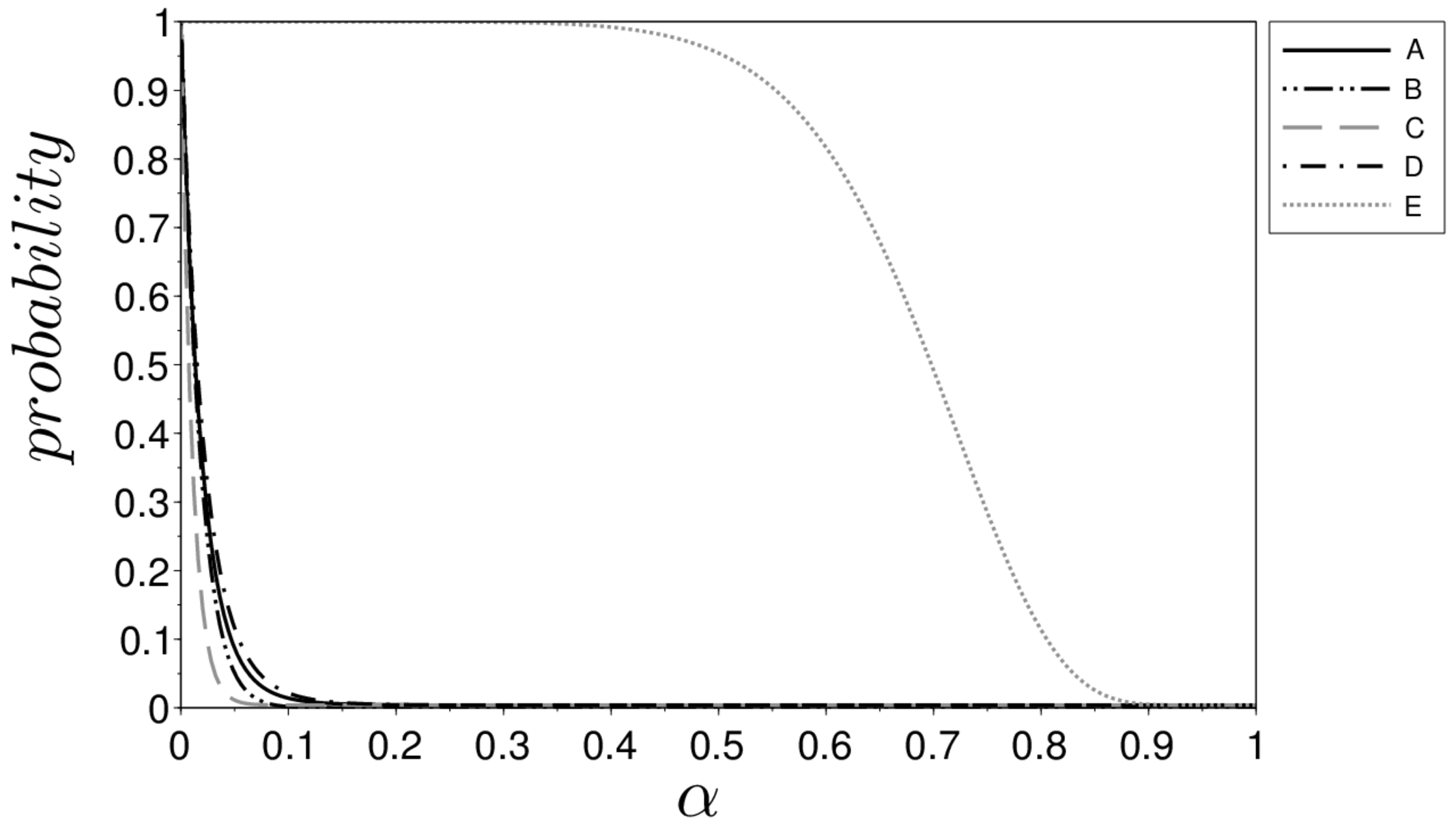}}
	\subfigure[Detail for small values of $\alpha$.]{
	\label{cotas_nuevas_pk_vs_alfa_kGr_NFijo_zoom} 
	\includegraphics[width=0.65\textwidth,keepaspectratio=true]
	{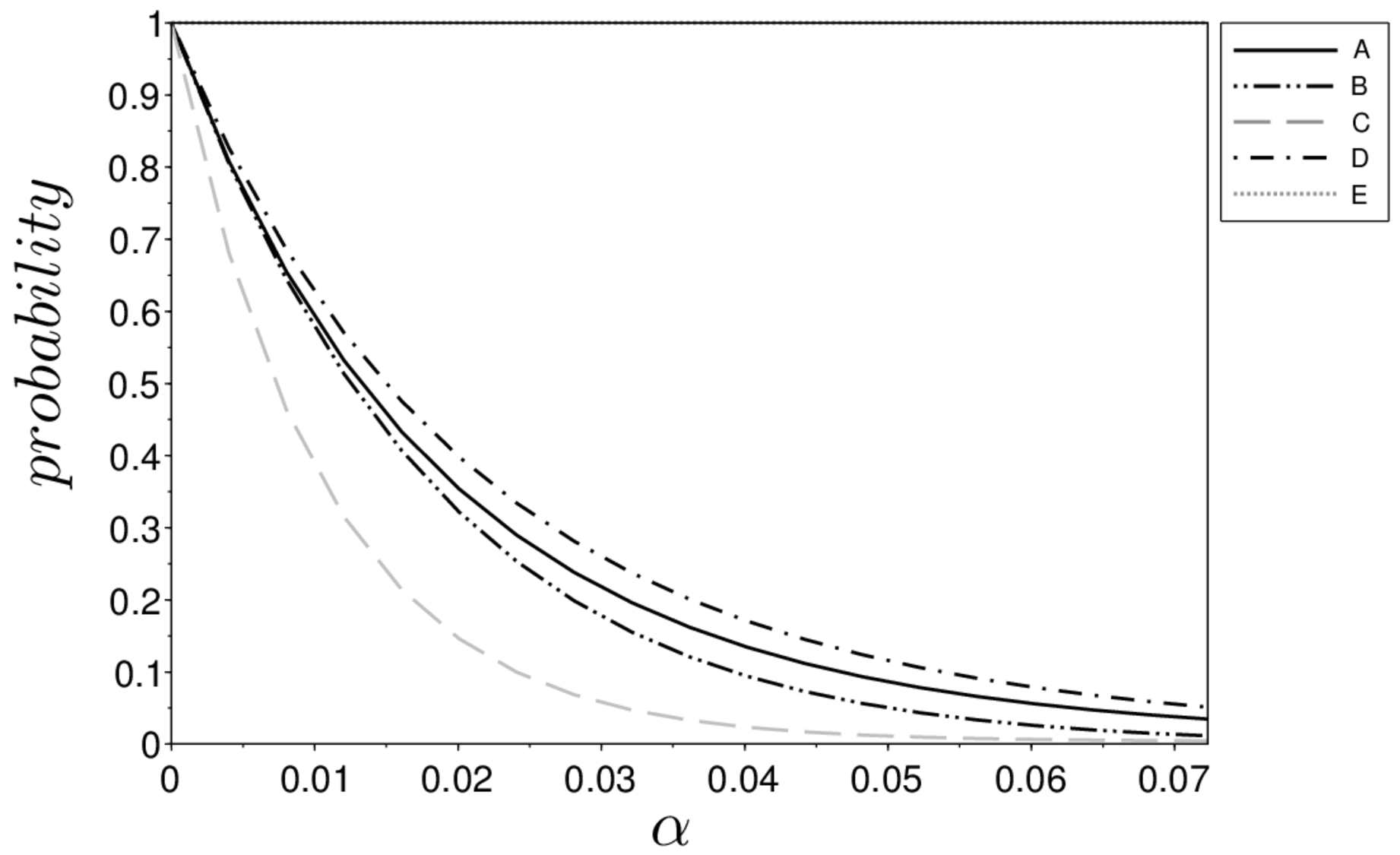}}
	\caption{Comparison between different bounds of probabilities: numerical simulation (A), first order approximation (B), lower bound with $\gamma_l(\alpha) = 1 - (1-\alpha)^n$ (C), upper bound with $\gamma_u(\alpha)=n\alpha/(2 + n\alpha)$ (D) and upper bound with $\gamma_u(\alpha) = \alpha^n$ (E).}
	\label{cotas_nuevas_pk_vs_alfa_kGr_NFijo_FIGURA_ENTERA} 
\end{figure}

In order to find upper and lower probability bounds, we assume the best and worst case for the desired probability, by taking $\rhoun \equiv G \rho G^{\dagger}$ or $\rhoun \equiv I/N$, respectively. Thus, the probability bounds are
\begin{eqnarray}
	\hat{p}_{u}(k, \alpha) & = & \left(1-\alpha^n\right)^k p(k) + \frac{1-\left(1-\alpha^n\right)^k}{N} \label{sec_app_cota_probs_u} \\
	\hat{p}_{l}(k, \alpha) & = &\left(1-\alpha\right)^{nk} p(k) + \frac{1-\left(1-\alpha\right)^{nk}}{N}, \label{sec_app_cota_probs_l}
\end{eqnarray}
that hold
\begin{equation}
\label{sec_app_cota_probs_relaciones_cota}
	\hat{p}_{l}(k, \alpha) \leq \hat{p}^L(k, \alpha) \leq \hat{p}_{u}(k, \alpha),
\end{equation}
$\forall k \in \mathbb{N},\alpha \in [0,1]$. Since these are of the form of the TDCh (Eq. (\ref{eq_rho_k_techo_total})), we define these bounds as their equivalent width of error $\gamma(\alpha)$, i.e.
\begin{eqnarray}
  \gamma_{u}(\alpha) & = & \alpha^n, \label{sec_app_cota_gammas_u} \\
	\gamma_{l}(\alpha) & = & 1-\left( 1-\alpha \right)^n. \label{sec_app_cota_gammas_l}
\end{eqnarray}
Even though $\gamma_{l}(\alpha)$ provides a reasonable probability bound, $\gamma_{u}(\alpha)$ does not (see Fig. \ref{cotas_nuevas_pk_vs_alfa_kGr_NFijo_FIGURA_ENTERA}). A better approach is to replace it by 
\begin{equation}
\label{sec_app_cota_gamma_upper}
	\gamma_{u}(\alpha) = \frac{n\alpha}{2 + n\alpha}.
\end{equation}

\subsection{First order approximation of the LDCh}
\label{subsec_groverLDC_FOA}
Here we will propose a method for analyzing the effect of the LDCh in Grover's algorithm in terms of its first order approximation (in $\alpha$, as in \cite{Nielsen_2000}), proving also that this approximation serves both as an approximation for small values of width of error $\alpha$ and as a lower probability bound for any $\alpha$.

After applying the LDCh in every qubit and every step (Eq. (\ref{eq_ro_k_techo_LDCh})), the probability of success after $k$ steps is of the form
\begin{equation}
\label{sec_app_cota_forma_probs}
	\hat{p}^L(k, \alpha) = \sum_{i=0}^{nk}\left(1-\frac{3\alpha}{4}\right)^{nk-i}\left(\frac{\alpha}{4}\right)^i f_i(n, k),
\end{equation}
where $i$ is the total number of \emph{errors} up to step $k$.
Hence, the first order approximation is
\begin{equation}
	\hat{p}^L(k, \alpha) \approx \left(1-\frac{3\alpha}{4}\right)^{nk}f_0(n, k) + 
	\left(1-\frac{3\alpha}{4}\right)^{nk-1}\left(\frac{\alpha}{4}\right) f_1(n, k),
\label{sec_app_cota_approx}
\end{equation}
where $f_0(n, k) \equiv p(k)$ is Grover's algorithm original probability and $f_1(n, k)$ corresponds to the all the contributions of applying \emph{one} error in any qubit at any step of the algorithm (see \ref{App_FindingF1}). As can be seen in Fig. \ref{cotas_nuevas_pk_vs_alfa_kGr_NFijo_FIGURA_ENTERA}, Eq. (\ref{sec_app_cota_approx}) is also a lower bound of the probability because all $f_i(n, k)$ are non-negative (sums of probabilities).

%
%
\subsection{Order estimation of the algorithm}
\label{subsec_groverLDC_orden}

Here we denote the estimate mean cost of the algorithm with LDCh error by $MC^L_{k, \alpha} \coloneqq MC^L\left[ k, \hat{p}^L(k, \alpha) \right]$. Using the monotonicity of both $\hat{p}^L(k, \alpha)$ and $(\gamma_l - \gamma_u)$ in $\alpha$, and Eq. (\ref{sec_app_cota_probs_relaciones_cota}), we have found bounds for the mean cost as follows:
\begin{equation} \label{eq_cotas_k_fijo_ordenes}
	MC_{k, \gamma_u} \leq MC^L_{k, \alpha} \leq MC_{k, \gamma_l}.
\end{equation}
That is, given a fixed number of steps $k$, the bounds $\hat{p}(k, \gamma_u)$ and $\hat{p}(k, \gamma_l)$ induce bounds on the mean cost.

%
%
\subsubsection{Stopping the algorithm at $k_{Gr}$}
\label{subsubsec_groverLDC_orden_kGr}
Considering $k_{Gr}$ as the final step, we have an estimate mean cost of the algorithm given by $MC^L_{k_{Gr}, \alpha}$, bounded as in Eq. (\ref{eq_cotas_k_fijo_ordenes}). Similar to the TDCh case, for any fixed $\alpha > 0$ the order is $\Theta(N \sqrt{N})$.

For $N \gg 1$ and small values of $\alpha$ ($\alpha k_{Gr}\log_2{N} \ll 1$) the mean cost becomes
\begin{equation} \label{eq_cotas_ordenes_kgr}
	 MC^L_{k_{Gr}, \alpha} = \frac{\pi \sqrt{N}}{4} + \zeta \alpha N \pi^2 \log_2{N},
\end{equation}
where $\zeta$ is a constant in the interval $\left[ 1/32, 1/16 \right]$ (from bounds). Using the first order approximation, Eq. (\ref{sec_app_cota_approx}) and $f_1(n,k)$, we have found that $\zeta$ is approximately $70/2048$. The latter is much closer to $1/32$ than to $1/16$, and hence, the actual probability is much closer when using $\gamma_u$ than when using $\gamma_l$. Because of the simplicity and the validity of the expression, we encourage to use $\gamma_u$ to estimate the probability $\hat{p}^L(k, \alpha)$.

From Eq. (\ref{eq_cotas_ordenes_kgr}), one can see that to mantain the quadratic speedup $\alpha \ll 2/(k_{Gr}\log_2{N})$ is needed.

%
%

 \subsubsection{Stopping the algorithm at $k_\emph{max}$}
 \label{subsubsec_groverLDC_orden_kMax}
We now consider the induced bounds for $\kmax$, which are justified by the continuity of $\hat{p}(k, \gamma)$ in both $\gamma$ and $n$. These are
\begin{equation} \label{eq_alfa_cotas_k_max}
	 k_{\textrm{max}\_\,l} \leq \kmax^L \leq k_{\textrm{max}\_u},
\end{equation}
where $k_{\textrm{max}\_\,l} = \kmax(\gamma_l)$ and $k_{\textrm{max}\_u} = \kmax(\gamma_u)$. Since we are stopping at the step of maximum probability (in each curve), the following bound holds:
\begin{equation} \label{eq_alfa_ordenes_cotas}
	MC_{k_{\textrm{max}\_\,l}, \gamma_u} \leq MC^L_{\kmax^L, \alpha} \leq MC_{ k_{\textrm{max}\_u}, \gamma_l}.
\end{equation}
In fact, using Eqs. (\ref{eq_alfa_ordenes_cotas}) and (\ref{eq_p1_alpha_n_grande}), one finds that $\kmax^L \rightarrow 1$ and the mean cost is given by $1/p^L(1, \alpha)$. Therefore, again, for any fixed $\alpha > 0$ the order is $\Theta(N)$. 

For $N \gg 1$ and small values of $\alpha$ ($ 0.55\alpha N \log_2{N} \ll 1$), the mean cost is approximately 
\begin{equation} 
\label{eq_MCL_bounds}
	MC^L_{\kmax^L, \alpha} = \frac{\pi \sqrt{N}}{4} + \zeta \alpha N \log_2{N},
\end{equation}
where $\zeta$ is a constant in the interval $\left[\left(\pi^2-4\right)/32,\left(\pi^2-1\right)/16\right]$. And for the algorithm to be of order $\Theta(\sqrt{N})$, one needs $\alpha \ll 2.5/(k_{Gr}\log_2{N})$.

This is coherent with the results presented with the TDCh model, where for small 
values of $\gamma$, choosing $k_{Gr}$ or $\kmax$ does not change the order 
of the algorithm, but for any fixed $\gamma > 0$, it changes drastically. 

Analogously to the work done in Sec. \ref{sec_groverTDC}, one can find 
$\alphac$: the maximum value of $\alpha$ that leaves the mean cost 
$N/2$. Considering the worst case yields
 \begin{equation} 
 \label{eq_alfa_orden_clasico}
 	\alphac \leq \frac{14}{\log_2{N}}.
\end{equation}
In contrast to the TDCh case, this width of error does depend on $N$.

\section{Conclusions}
\label{sec_conclusions}
In this article we have studied the effect of noise in Grover's quantum search algorithm, based in two models of error: total depolarizing channel (TDCh) and local depolarizing channel (LDCh). Our focus was not in error correction (e.g. by the fault-tolerant method), but to provide an insight to the kind of error, or degradation, that needs to be corrected. Our work extends the results found by Vrana et. al. regarding only the TDCh model,  concluding that the LDCh error model is more devastating than the TDCh error. We also show that, for both cases, with an $N$-dependent small enough error-width, correction is not needed.

The devastating effect of the LDCh was, in some sense, expected: in the case of TDCh model the state is partially altered, but the rest remains intact (the state evolves partially in the plane formed by $\ket{s} $ and $\ket{t}$, or otherwise becomes $I/N$). Whereas, with an LDCh error model, the state is mostly removed from the plane formed by $\ket{s}$ and $\ket{t}$.

We have found a closed form for the step at which the maximum of probability occurs for the TDCh error model ($\kmax$), and provided bounds for the LDCh error model (based on the TDCh model). This step could be used as an indicator of the impact of the error in the algorithm. This means, comparing $\kmax$ with $k_{Gr}$ one has an idea of how degraded the algorithm is. Analogous comparisons could be done with the probability at such steps, but are, in some sense, equivalent. 

From this study we conclude that to maintain the original speed up of Grover's algorithm, different requirements are needed depending on the error model: in the TDCh case, the error-width has to be smaller than $1/\sqrt{N}$ (as in \cite{Vrana_2014}); and for the LDCh model, it needs to be smaller than $1/(\sqrt{N}\log_2{N})$.

Furthermore, there are widths of error for both the TDCh and the LDCh model for which 
the algorithm still outperforms the classical brute-force algorithm. It is interesting to note 
that for the TDCh model this width does not depend on $N$, i.e. for widths below $7/8$ the algorithm is better than the classical (Eq. (\ref {eq_orden_kmax_gama_grande})). Nevertheless, for the LDCh model, this classical threshold error $\alphac$, depends of $N$ (Eq. (\ref{eq_alfa_orden_clasico})) and goes to zero with $N$, reinforcing the idea that the LDCh degrades more the algorithm than the TDCh.

The depolarizing channel is a particular error model. It would be interesting to study other error models taking into account experimental implementations of the algorithm. Future work should be done with this focus, and possibly including error correction.

\appendix

\section{Finding $\kmax$}
\label{sec_appendix_subsec_finding_kmax}

We wish to find $\kmax$ which maximizes Eq. (\ref{eq_p_k_techo_total}). We will assume $k \in \mathbb{R}$, and then perform some adjustments. Considering $p(k)$ as in Eq. (\ref{eq_p_k}), we find the first derivative of Eq. (\ref{eq_p_k_techo_total})
\begin{equation}
\frac{d \hat{p}(k, \gamma)}{dk} = \left( 1-\gamma \right)^k \left[  2\theta \sin{(2(2k+1)\theta)} + \label{eq_apendice_p_k_techo_total_derivada} 
\left(p(k) - \frac{1}{N} \right)\ln{(1-\gamma)} \right]. 
\end{equation}

Assuming $\gamma \neq 1$ and nulling the derivative yields
\begin{equation} \label{eq_apendice_p_k_techo_total_derivada_cero_0}
4\theta \sin{((2k+1)\theta)}\cos{((2k+1)\theta)} 
+ \left(\sin^2{((2k+1)\theta)} - \frac{1}{N}\right)\ln{(1-\gamma)} = 0.
\end{equation}

The former can be expressed as
\begin{equation} \label{eq_apendice_p_k_techo_total_derivada_cero_1}
	-C = -A\sin^2{x} + B\sin{x}\cos{x},
 \end{equation}
 where $x = (2k+1)\theta$, $A = -\ln{(1-\gamma)}$, $B = 4\theta$ and $C = -\ln{(1-\gamma)}/N$. This can be reduced to 
\begin{equation} \label{eq_apendice_p_k_techo_total_derivada_cero_2}
	-z = \sin{(x)} \cos{(x + \phi)},
 \end{equation}
where $z = C / \sqrt{A^2+B^2}$ and $\phi = \arcsin{\left( A / \sqrt{A^2+B^2}\right)}$. Considering $x>0$ Eq. (\ref{eq_apendice_p_k_techo_total_derivada_cero_2}) has two family of solutions
 \begin{eqnarray} \label{eq_apendice_k_max_familias}
	x_{1} & = & \pi m - \frac{1}{2} \arcsin(2 z-\sin (\phi ))-\frac{\phi }{2}, \quad \text{and} \nonumber \\
	x_2 &=& \frac{1}{2} \left(2 \pi m+\arcsin(2 z-\sin (\phi ))-\phi +\pi \right), 
\end{eqnarray}
being $m$ an arbitrary integer. Thus, the solution to the original problem is
\begin{equation} \label{eq_apendice_k_max_antes_ajustes}
	 \frac{\pi - 2\theta - \arcsin{(\delta)}  - \arcsin{\left( \left[1-\frac{2}{N} \right]\delta \right)}}{4\theta}.
\end{equation}
where
\begin{equation} \label{eq_apendice_k_max_familias_parte}
\delta = \sqrt{\frac{1}{1 + {\left(\frac{4 \theta}{\ln{(1-\gamma)}}\right)} ^2}}.
\end{equation}

Adjusting border values with $\gamma \rightarrow 0$ and $\gamma \rightarrow 1$, and taking integer part yields
\begin{equation}\label{eq_apendice_k_max_fin}
	\kmax(\gamma) = \max\left( \left\lfloor  \frac{\pi - \arcsin{\delta} - \arcsin{\left(  \left[ 1-\frac{2}{N} \right] \delta \right) }}{4\theta}  \right\rfloor , 1\right).
\end{equation}

\section{Finding $f_1(n, k)$}
\label{App_FindingF1}
As stated before, $f_1(n, k)$ corresponds to the depolarizing channel acting in any qubit at any step of the algorithm. Because of the simmetry of Grover's algorithm in the qubits, one can assume without loss of generality that the last qubit is the affected one and then multiply by the number of possible ways of having the depolarizing channel acting on one qubit ($n$). We can also assume, without loss of generality, that the target state is $\ket{t}\equiv \ket{00\cdots0} = \ket{0}$.

The depolarizing channel acting on one qubit (without the identity operator and considering that error has been commited) can be expressed with the operators
\begin{equation}
X \coloneqq I^{\otimes n-1}\otimes \sigma_x, \ Y \coloneqq I^{\otimes n-1}\otimes \sigma_y, \
\ Z  \coloneqq  I^{\otimes n-1}\otimes \sigma_z, \label{sec_app_cota_subsec_f1_operadores}
\end{equation}
where $I$ represents the $1$-qubit identity operator and $\sigma_\mu$ the Pauli matrices.

It is useful to see that the set $\{\ket{0}, \ket{s}, \ket{1}, \ket{p}\}$ is closed under the operations of $\{X, Y, Z, G\}$, where $G$ is Grover's operator and $\ket{p}$ is defined as the normalized superposition of the even states of the canonical base as 
\begin{equation}
\label{sec_app_cota_subsec_f1_definicion_p}
	\ket{p}  \coloneqq  \sqrt{\frac{2}{N}}\sum_{i=0}^{\frac{N}{2}-1}\ket{2i}.
\end{equation}

The state obtained after $k$ steps is represented as $\phi(k) = \left[a_k \quad b_k \quad c_k \quad d_k\right]^T$, which corresponds to the quantum state $\ket{\varphi_k} = a_k\ket{0} + b_k\ket{s} + c_k\ket{1} + d_k\ket{p}$. The probability of obtaining the marked state ($\ket{0}$) is given by $\left[ tr\left(C.\phi(k) \right) \right]^2$, where $C = \left[1 \quad \lambda/2 \quad 0 \quad \lambda/\sqrt{2} \right]$ and $\lambda = 2/\sqrt{N}$.
Hence, the desired function $f_1(n, k)$ is
\begin{equation}
\label{sec_app_cota_subsec_f1_probabilidades_suma}
	f_1(n, k) = \sum_{l=1}^{k}\sum_{j=1}^{3}\left[ tr\left(C A^{k-l} B_{j} A^{l} \phi(0) \right) \right]^2,
\end{equation}
where $\phi(k) = \left[0 \quad 1 \quad 0 \quad 0 \right]^T$ is the initial step of the algorithm ($\ket{s}$) and $A, B_1, B_2, B_3$ correspond to applying $G, X, Y,$ or $Z$ respectively (up to a global phase). Such matrices are given by
\begin{equation}
A =  \left[\begin{array}{cccc}
		-1 & -\lambda & 0 & -\sqrt{2}\lambda\\
		\lambda & \lambda^{2}-1 & -\lambda & \sqrt{2}(\lambda^{2}-1)\\
		0 & 0 & 1 & 0\\
		0 & 0 & 0 & 1
	\end{array}\right], \
B_1 = \left[\begin{array}{cccc}
		0 & 0 & 1 & 0\\
		0 & 1 & 0 & \sqrt{2}\\
		1 & 0 & 0 & 0\\
		0 & 0 & 0 & -1
	\end{array}\right],
\end{equation}

\begin{equation}
B_2 = \left[\begin{array}{cccc}
		0 & 0 & -1 & 0\\
		0 & 1 & 0 & \sqrt{2}\\
		1 & 0 & 0 & 0\\
		0 & -\sqrt{2} & 0 & -1
	\end{array}\right] \ \textrm{and} \
B_3 = \left[\begin{array}{cccc}
		1 & 0 & 0 & 0\\
		0 & -1 & 0 & 0\\
		0 & 0 & -1 & 0\\
		0 & -\sqrt{2} & 0 & 1
	\end{array}\right].
\end{equation}

The expression of $f_1(n,k)$, derived here, has been used to perform the series expansion made in Eq. (\ref{eq_cotas_ordenes_kgr}). The full expression is too large to include in this article.

\bibliographystyle{unsrt}
\bibliography{GroverDepChn_2015}

\end{document}